\documentclass[sigconf]{acmart}

\AtBeginDocument{%
  \providecommand\BibTeX{{%
    \normalfont B\kern-0.5em{\scshape i\kern-0.25em b}\kern-0.8em\TeX}}}

\usepackage{balance}
\usepackage{booktabs}
\usepackage{amsmath}
\usepackage{amsfonts}
\usepackage{comment}
\usepackage{footmisc}
\usepackage{mathrsfs}
\usepackage{graphicx}
\usepackage{subfigure}
\usepackage{multirow}
\usepackage{algorithm}
\usepackage{algorithmic}
\usepackage{multicol}
\usepackage{enumitem}
\usepackage{array}
\usepackage{float}
\usepackage{hyperref}
\usepackage[title]{appendix}

\usepackage{xspace}

\newcommand{\ie}{\emph{i.e.,}\xspace}
\newcommand{\name}{GFN4Retention\xspace}

\copyrightyear{2024}
\acmYear{2024}
\setcopyright{acmlicensed}\acmConference[KDD '24]{Proceedings of the 30th ACM SIGKDD Conference on Knowledge Discovery and Data Mining}{August 25--29, 2024}{Barcelona, Spain}
\acmBooktitle{Proceedings of the 30th ACM SIGKDD Conference on Knowledge Discovery and Data Mining (KDD '24), August 25--29, 2024, Barcelona, Spain} \acmDOI{10.1145/3637528.3671531} \acmISBN{979-8-4007-0490-1/24/08}

\begin{document}

\title{Modeling User Retention through Generative Flow Networks}

\author{Ziru Liu\textsuperscript{\dag}}
\affiliation{%
\institution{City University of Hong Kong}
\city{Hong Kong}
\country{China}}
\email{ziruliu2-c@my.cityu.edu.hk}

\author{Shuchang Liu\textsuperscript{\dag}}
\affiliation{%
  \institution{Kuaishou Technology}
  \city{Beijing}
  \country{China}
  }
\email{liushuchang@kuaishou.com}

\author{Bin Yang}
\affiliation{%
  \institution{Kuaishou Technology}
  \city{Beijing}
  \country{China}
  }
\email{yangbin11@kuaishou.com}

\author{Zhenghai Xue}
\affiliation{%
  \institution{Nanyang Technology University}
  \city{Singapore}
  \country{Singapore}
  }
\email{zhenghai001@e.ntu.edu.sg}

\author{Qingpeng Cai*}
\affiliation{%
  \institution{Kuaishou Technology}
  \city{Beijing}
  \country{China}
  }
\email{caiqingpeng@kuaishou.com}

\author{Xiangyu Zhao*}
\affiliation{%
\institution{City University of Hong Kong}
\city{Hong Kong}
\country{China}}
\email{xianzhao@cityu.edu.hk}

\author{Zijian Zhang}
\affiliation{%
\institution{City University of Hong Kong}
\city{Hong Kong}
\country{China}}
\email{zijian.zhang@my.cityu.edu.hk}

\author{Lantao Hu}
\affiliation{%
  \institution{Kuaishou Technology}
  \city{Beijing}
  \country{China}
  }
\email{hulantao@kuaishou.com}

 \author{Han Li}
 \affiliation{%
   \institution{Kuaishou Technology}
   \city{Beijing}
   \country{China}
   }
 \email{lihan08@kuaishou.com}

\author{Peng Jiang*}
\affiliation{%
  \institution{Kuaishou Technology}
  \city{Beijing}
  \country{China}
  }
\email{jiangpeng@kuaishou.com}

\thanks{\dag Co-first authors. * Corresponding authors.}

\renewcommand{\shortauthors}{Ziru Liu et al.}

\begin{abstract}
Recommender systems aim to fulfill the user's daily demands.
While most existing research focuses on maximizing the user's engagement with the system, it has recently been pointed out that how frequently the users come back for the service also reflects the quality and stability of recommendations.
However, optimizing this user retention behavior is non-trivial and poses several challenges including the intractable leave-and-return user activities, the sparse and delayed signal, and the uncertain relations between users' retention and their immediate feedback towards each item in the recommendation list.
In this work, we regard the retention signal as an overall estimation of the user's end-of-session satisfaction and propose to estimate this signal through a probabilistic flow.
This flow-based modeling technique can back-propagate the retention reward towards each recommended item in the user session, and we show that the flow combined with traditional learning-to-rank objectives eventually optimizes a non-discounted cumulative reward for both immediate user feedback and user retention.
We verify the effectiveness of our method through both offline empirical studies on two public datasets and online A/B tests in an industrial platform. The source code is accessible to facilitate replication \footnote{\url{https://github.com/Applied-Machine-Learning-Lab/GFN4Retention}}.
\end{abstract}



\maketitle

\section{Introduction}
In the era of information abundance, recommender systems have become essential tools that guide users to content that resonates with their personal preferences ~\cite{aceto2020industry}. 
Traditional metrics used to evaluate these systems — such as clicks, likes, and ratings — are adept at capturing user preferences for each recommended item \cite{das2007google,zhao2019deep} and are formulated as targets to guide the optimization of the recommender systems.
Despite their effectiveness, they essentially estimate the user's immediate feedback of items and are incapable of providing a comprehensive assessment of users' long-term engagement ~\cite{wu2017returning,yi2014beyond}. 
For example, 
when the system finds that items with compelling features (e.g. addictive content) can maximize the click rate, it may decide to continuously recommend such items.
However, these features may initially be appealing but quickly lose the user's fondness. 
This discrepancy suggests a gap between the user's immediate interest in an item and the sustainable interest ~\cite{viljanen2016modelling} of the system.
As a resolution, long-term metrics are adopted to offer deeper insights into users' overall satisfaction.
One typical example is the user retention signal that describes the user's return-to-app behavior.
This metric is one of the most critical performance estimators for many online services since it closely correlates with the pivotal business indicator, i.e. Daily Active Users (DAU)~\cite{zou2019reinforcement}.

In practice, modeling and optimizing user retention is a challenging task because of its between-session nature.
Specifically, the retention behavior does not happen until the user leaves the current session and returns at the beginning of the next session.
And it has no clear relation to any single recommendation step in previous interactions.
Furthermore, the user's activity between the two consecutive sessions is assumed unobservable for the service, introducing extra uncertainty.
To circumvent these challenges, evidence has found promising results using reinforcement learning (RL) to optimize the cumulative reward for the entire user interaction sequence~\cite{chen2021survey,cai2023reinforcing} as a surrogate.
Intuitively, the user returns to the platform because the system's overall impression is sufficiently positive and attractive, which can be partially measured by the sum of (positive feedback related) rewards in the session.
RL-based recommendation solutions solve the optimization of this cumulative reward by formulating the user interaction sequence as a Markov Decision Process (MDP) and learning a policy that considers the long-term impact of each recommendation action~\cite{zhao2011reinforcement,afsar2021reinforcement}. 
This allows them to dynamically tailor recommendations at each point of user interaction, adapting to the evolving preferences and optimizing the cumulative reward of the whole session.

Yet, the relation between the session-level cumulative reward and user retention is still unclear, so the authors in~\cite{cai2023reinforcing} further show that the RL-based method may directly integrate the cross-session retention signal into the long-term value estimation.
Though effective, this method is not designed to investigate the influence of each interaction on the retention signal (denoted as ``retention attribution'') and it also indirectly optimizes the retention with cumulative immediate rewards as a surrogate.
Besides, all RL-based solutions may suffer from the exploration and exploitation trade-off~\cite{ciosek2019better} that limits their performance on unstable metrics.
This instability is particularly pronounced in scenarios where user retention dynamics are complex, uncertain, and rapidly evolving.
In general, we want to have a stable exploratory solution that can simultaneously optimize user retention and immediate rewards.

Inspired by the recent development of Generative Flow Networks (GFNs)\cite{bengio2021flow,zhang2023distributional,pan2023pre,pan2023better,pan2023stochastic,pan2022generative,zhang2023distributional,zhang2024let,lau2024qgfn}, we propose an alternative approach \textbf{\name} that considers the session-level recommendation as a generation task where the retention signal is directly modeled by the trajectory generation probability.
Similar to the general GFN formulation, a probabilistic generation process will finally construct a user session (trajectory) where the target retention reward is matched only by the end of the process.
Specifically, each recommendation step is considered as a conditional forward probabilistic flow to the next user state and each user state is associated with a flow estimator that represents the generation probability of reaching this state.
During optimization, the end-of-session terminal state directly matches the generation probability with the retention reward.
For non-terminal states, a flow matching learning objective that incorporates an additional backward probabilistic flow is used to back-propagate the retention reward towards every step in the sequence.
This would implicitly model each recommendation action's retention attribution.
As discussed in~\cite{bengio2023gflownet}, GFNs have demonstrated remarkable strength in generating diverse objects of high quality, which naturally solves the aforementioned exploration challenge of RL.
Nevertheless, incorporating GFNs in the recommendation task with retention optimization also brings new challenges.
In its design, GFNs may find it difficult to track the nuanced changes in user engagement of each recommendation step, since it originally assumes the absence of intermediate rewards.
As a countermeasure, we derive an integrated reward system that can balance the immediate reward and the retention attribution in each recommendation step.
The final reward of \name derives a refined detailed balance loss that matches the session-level generation probability with the product of the retention reward and the non-discounted cumulative reward of immediate feedback.
Secondly, the recommendation space is combinatorially large and the list-wise recommendation is generated and represented by a point in the continuous vector space~\cite{cai2023reinforcing,liu2023hac,xue2022resact} in practice. We show that the flow matching property still holds in this continuous space and illustrate how to design each flow estimation component accordingly.
To this end, we summarize our key contributions in this paper as follows:
\begin{itemize}[leftmargin=*]
    \item We emphasize the importance of retention optimization in real-world RS applications and introduce \name learning framework designed for enhancing user retention while maintaining good exploration.
    \item Our proposed solution innovatively derived an integrated reward design with a refined detailed balance objective that can control the trade-off between the immediate reward and the retention attribution in each recommendation step. Besides, we also propose to optimize the flow-matching objectives in the continuous action space to accommodate the large recommendation space of the item list.
    \item We validate the superiority of \name through extensive experiments compared with state-of-the-art RL-based recommendation models on both offline and live experiments, and discuss the behaviors of each component with ablation study and parameter analysis.
\end{itemize}
\section{Preliminaries}
This section provides an overview of Generative Flow Networks and delves into the specific problem in our study: the session-wise recommendation for retention and immediate reward optimization.

\begin{figure*}
    \centering    \includegraphics[width=0.75\linewidth]{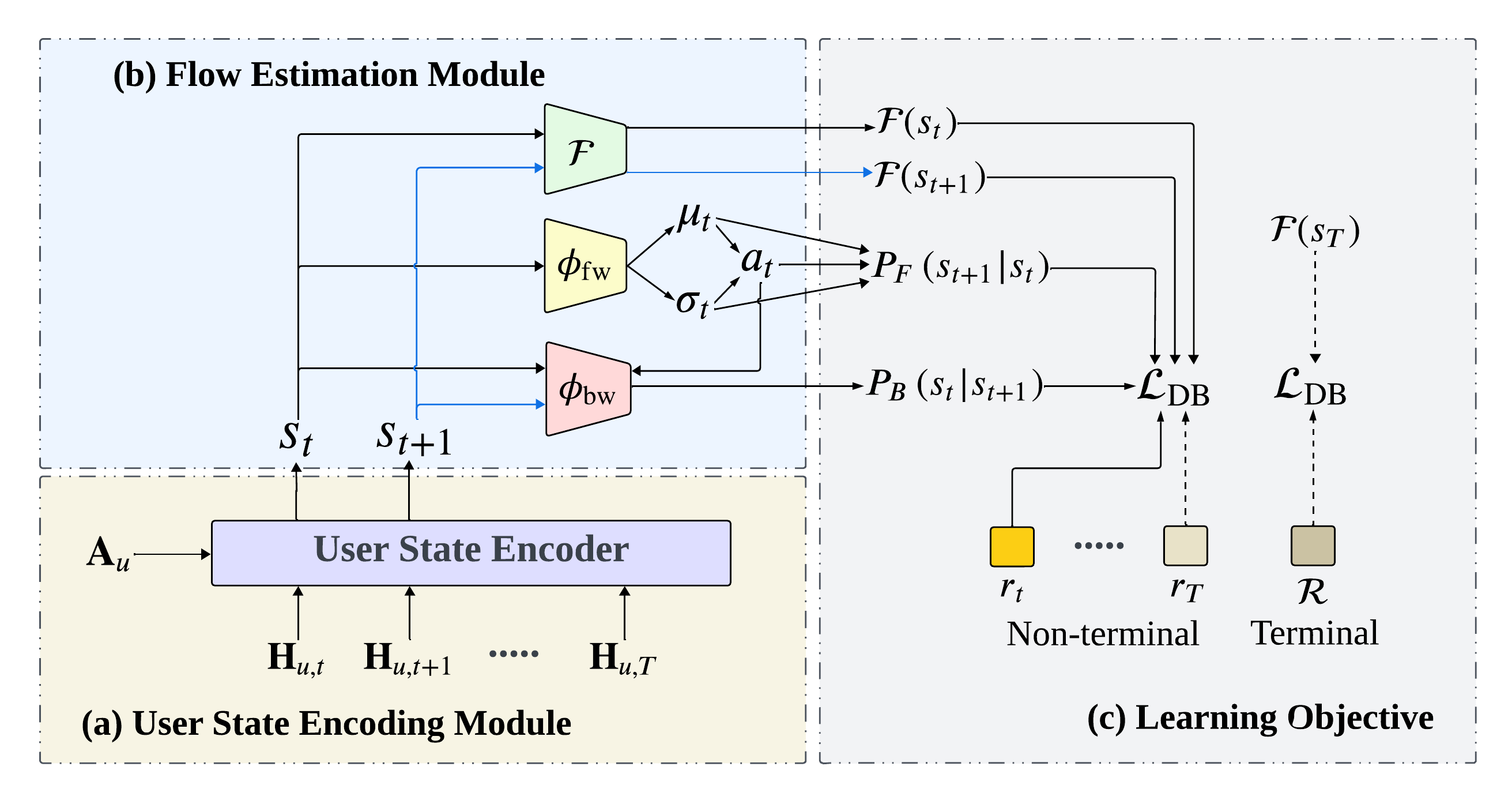}
    \vspace*{-4mm}
    \caption{Overview of GFN4Retention Framework}
    \label{fig:framework}
    \vspace*{-4mm}
\end{figure*}

\subsection{Generative Flow Networks}
In recent years, Generative Flow Networks (GFNs) have emerged as an innovative force in generative modeling, introducing a methodology for learning and sampling within complex distributions. 
The core of GFNs' design is its ability to develop a generation policy that directs the probabilistic flow across a state space towards terminal states \cite{bengio2021flow, bengio2023gflownet}. 
Specifically, the general formulation considers a generation trajectory $\mathbf{S}=\left\{s_1 \rightarrow s_2 \rightarrow \cdots \rightarrow s_T\right\}$ and assumes a forward probabilistic policy $P_F(s_{t+1}|s_t)$ that determines each action during the generation process.
Here $s_t$ represents the state at step $t$, and $T$ is the terminal step of the trajectory.
The goal of GFNs is to align the generation probability $P(\mathbf{S})$ of each trajectory with the observed reward by the end:
\begin{equation} \label{eq1}
    P(\mathbf{S}) \propto R(s_T)
\end{equation}
which is essentially an energy-based method since the model simultaneously serves as the generator and evaluator.



\noindent During optimization, the GFN framework introduces a flow estimator \( \mathcal{F}(s_t) \) that assesses the likelihood of traversing through a specific state $s_t$. The learning process of flow in GFNs is meticulously calibrated to align with the target distribution, ensuring equilibrium between the sums of incoming and outgoing flows:
\begin{equation}
     \mathcal{F}\left(s_t\right) P_F\left(s_{t+1} \mid s_t  \right) \approx \mathcal{F}\left(s_{t+1}\right) P_B\left(s_t \mid s_{t+1}  \right)
\end{equation}
where $P_F\left(s_{t+1} \mid s_t  \right)$ is the forward probability from $s_t$ to $s_{t+1}$, and $P_B\left(s_t \mid s_{t+1}  \right)$ is the corresponding backward probability that models the likelihood of the source state $s_t$ given the outcome state $s_{t+1}$.

The foundational work on GFNs has led to the development of an optimized variant for this objective: the Detailed Balance (DB) loss ~\cite{bengio2023gflownet} that minimizes the difference between the forward view and backward view of each step-wise joint probability $P(s_t,s_{t+1})$. Then the objective follows the flow matching property in the generation process which minimizes the Detailed Balance (DB) loss:
\begin{equation} \label{eq:DB1}
\min \mathcal{L}_{\mathrm{DB}}\left(s_t, s_{t+1}\right)=\left(\log \frac{\mathcal{F}\left(s_t\right) P_F\left(s_{t+1} \mid s_t  \right)}{\mathcal{F}\left(s_{t+1}\right) P_B\left(s_t \mid s_{t+1}  \right)}\right)^2
\end{equation}
By minimizing the DB loss, it strives to achieve a flow representation that is highly indicative of the target distribution.

\subsection{Problem Definition}
\label{section:PD}
\begin{figure} [h]
    \centering    \includegraphics[width=0.98\linewidth]{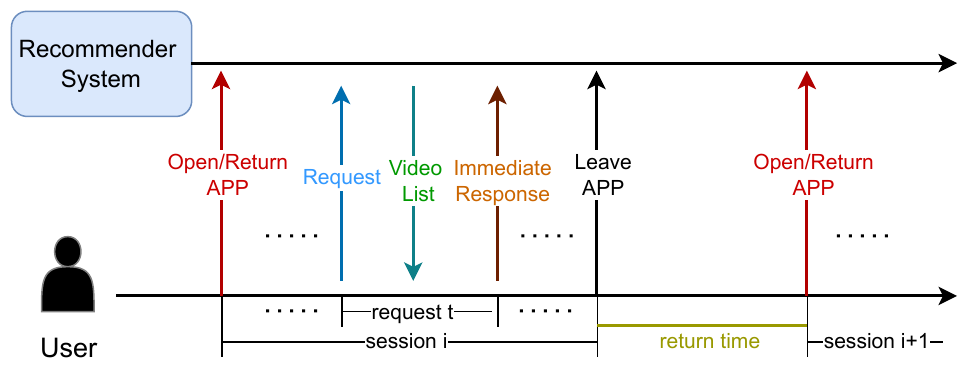}
    \vspace*{-4mm}
    \caption{Session-wise Recommendation Example}
    \label{fig:sessionrec}
    \vspace*{-4mm}
\end{figure}

In our research, we focus on the problem of session-wise recommendation, which aims to iteratively suggest items to users and maximize both retention and immediate feedback of a user session. The session-wise recommendation for a short video application scenario is illustrated in Figure \ref{fig:sessionrec}.
Formally, we consider a set of users \( \mathcal{U} \) and a set of items \( \mathcal{C} \). 
For each session, at any given step \( t \), we may receive a recommendation request from user \( u \in \mathcal{U} \), which consists of a user feature set $\textbf{A}_{u}$, the user's interaction history $\textbf{H}_{u,t}$.
In recommender systems, the user request provides the necessary context information to encode the current user \textbf{state} $s_t$.
Given the user request and the encoded state, the recommendation policy generates an \textbf{action} $a_t$ that corresponds to a list of items selected from $\mathcal{C}$.
Then the user provides feedback of $\mathcal{B}$ behavior types (e.g. clicks, likes, and comments) for these items which is used to calculate an \textbf{immediate reward} \( r_{t} \):
\begin{equation}
    r_t = \sum_{b \in \mathcal{B}} \omega_{b} \cdot y_{t,b}
\end{equation}
where \( y_{t,b} \) represents the user's feedback in behavior \( b \) at step $t$, and $\omega_{b}$ is the weight for behavior \( b \).
By the end of the session (i.e. at $s_T$), we also observe the \textbf{user retention reward} \( \mathcal{R} \) defined as the user return frequency which is the core metric in this work.
We organize each sample as the tuple $(\mathbf{S}, a_1,\dots,a_T,r_1,\dots,r_T, \mathcal{R})$ that consists of the observed states (i.e. user requests), actions, immediate rewards, and the retention reward of a session.
And we set two \textbf{goals} in our problem setting:
1) find a valid reward design $R(\mathbf{S})=f(r_1,\dots,r_T,\mathcal{R})$ that combines the retention reward and the cumulative immediate rewards, and helps boost the overall recommendation performance;
2) learn a recommendation policy that explores and achieves a better combined reward of $R(\mathbf{S})$.

\section{THE PROPOSED Framework}

In this section, we present our \name framework.
As illustrated in Figure \ref{fig:framework}, the main framework includes a feature extraction module that encodes user request context into the user state, a flow estimation module that models the generative flow of user sessions, and a modified detailed balance learning objective with integrated reward design.

\begin{figure} [h]
    \centering    \includegraphics[width=0.98\linewidth]{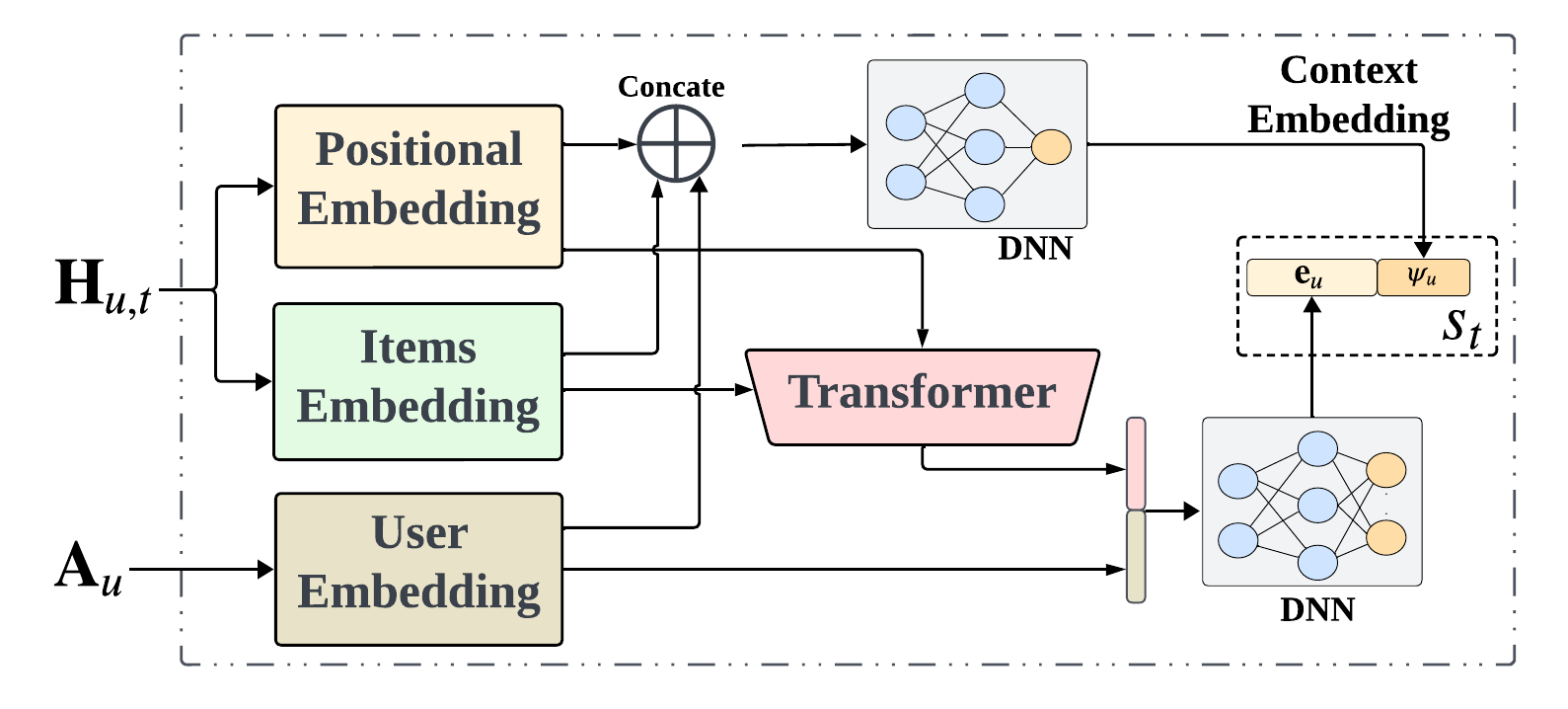}
    \vspace*{-4mm}
    \caption{Framework of User State Encoder}
    \label{fig:encoder}
    \vspace*{-4mm}
\end{figure}

\subsection{User State Encoding} \label{Section:encoder}

While real-world online platforms heavily depend on navigating the intricate user dynamics as well as their static features, the challenge of retention optimization in this scenario is not only the identification of diverse user patterns but also the agility to adjust to rapid changes in these patterns. In our design, upon receiving a recommendation request from a user $ u \in \mathcal{U} $, we consider two major types of input including the user's feature set $\mathbf{A}_{u}$ and the interaction history $\mathbf{H}_{u,t}$.
To better employ the dynamics in the user history and find patterns in the items' mutual influences, we first process the history $\mathbf{H}_{u,t}$ with a transformer, and the last output embedding is considered as the history encoding.
Then, a user feature embedding generated from $\mathbf{A}_u$ is concatenated with the history encoding and further processed by a neural network to generate an embedding $\mathbf{e}_u$ that forms the first part of the user state $s_t$.
In practice, we found that merely using a transformer to encode user history may over-amplify the most recent history and may ignore the feature-level interactions.
As a result, we include a DNN-based context-detecting module that encodes all the contexts in the user requests and outputs an addition embedding $\psi_u$ that forms the second part of the user state. 
The detailed framework of the feature encoder and context-detecting module is illustrated in Figure \ref{fig:encoder}, which details Part (a) of Figure \ref{fig:framework}.



\subsection{Recommendation Policy as Forward Flow}
During inference, with the encoded user state $s_t$, we can generate the action for the recommendation.
As widely adopted in many recommender systems, the service has to recommend a list of items for each user request to meet the latency requirement of users' frequent browsing behaviors.
This means that it is impractical to directly consider the enormous item set $\mathcal{C}$ as the action space.
Alternatively, we choose to consider a vector space for each action $a_t$, this vector can represent the output item list through a deterministic top-K selection module~\cite{liu2023hac}.
In our design, the policy network (i.e. the forward flow estimator) will first output the statistics of the Gaussian distribution $\mu,\sigma = \phi_\mathrm{fw}(s_t)$ and then sample the action vector as $a_t \sim \mathcal{N}(\mu,\sigma)$.
Note that this setting is mathematically different from the original design of GFNs where a small discrete action space is adopted.
Fortunately, the flow matching property in the continuous vector space still holds (explained in Appendix \ref{Appendix:continuous}) and we can safely apply the flow estimation and the detailed balance optimization as we will present in the following sections.
During training, the recommendation policy is regarded as the forward flow function $P_F(s_{t+1}|s_t)$ which assumes that the output action $a_t$ determines the next state $s_{t+1}$.


\subsection{Retention Flow Estimation}
Following the general design of GFNs, we include a flow estimator $\mathcal{F}(s_t)$ of states and a backward flow function $P_B(s_{t}|s_{t+1})$ that estimates the posterior.
In the session-level viewpoint, each observed session $\mathbf{S}$ is a probabilistic trajectory generated by the recommendation policy.
Each state $s_t$ may diverge into different future states according to the sampling process in $P_F$, and it may be reachable from various previous states.
Intuitively, the state flow estimator $\mathcal{F}(s_t)$ represents how likely a state $s_t$ is reached.
The backward function $P_B(s_t|s_{t+1})=\phi_\mathrm{bw}(s_t,a_t,s_{t+1})$ takes the current state-action and the next state as input and estimates how likely the next state is generated by the current state.
To guarantee the property $\mathcal{F}(s_t) \geq 0$ and $P_B(\cdot) \geq 0$, we use sigmoid activation for the network output.
Then, we can match the trajectory generation likelihood with the observed retention reward and use the flow matching objective (i.e. Eq\eqref{eq:DB1}) to back-propagate this end-of-session retention signal towards each intermediate step:
\begin{equation}
    \mathcal{L}_{\mathrm{DB}}= \begin{cases}\left(\log \frac{\mathcal{F}_{R}(s_t) \cdot P_F\left(s_{t+1} \mid s_t\right)}{\mathcal{F}_{R}(s_{t+1}) P_B\left(s_t \mid s_{t+1}\right)}\right)^2 &  1 \leq t \leq T - 1 \\ 
    \\
    \left(\log \frac{\mathcal{F}_{R}(s_t)}{\mathcal{R}}\right)^2 & t = T
\end{cases}
\end{equation}
where the forward function $P_F(s_{t+1}|s_t)$ depends on the recommendation policy $\phi_\mathrm{fw}$ as described in the previous section and we use $\mathcal{F}_R(s_t)$ to represents that the flow estimator in this framework only considers the retention reward.
As proven by~\cite{bengio2023gflownet}, this design ensures the learning goal $P(\mathbf{S})\propto\mathcal{F}_R(s_T) \approx \mathcal{R}$.
During the learning process, the generation may initially be random and generate low retention reward, but the exploration effect of the policy would gradually discover samples with higher retention and the actions in those sessions will receive higher generation possibility.
Eventually, this flow estimation learning framework helps the generation policy provide more diversity while maintaining high quality.

\subsection{Refined Detailed Balance Learning with Reward Integration}
Merely optimizing the retention as in the previous section only models the long-term preferences of users. 
In contrast, the immediate rewards in each intermediate step of the session provide valuable nuanced information that may be ignored by the long-term retention reward.
For example, a session that consists of a relevant item and an irrelevant item may still receive a good retention reward since the user may return as long as there is relevant information.
However, we should not consider the two items as equal and the user's feedback on each of the items may help us differentiate them.
Intuitively, we consider the retention reward and immediate rewards as complementary views of user preferences.
In this section, we illustrate how to integrate these two rewards in the generative flow estimation and derive a refined DB learning process.
\subsubsection{Reward Design:} 
To accommodate the flow estimation framework and guarantee the correctness of the detailed balance objective, we propose a reward integration through the product:
\begin{equation} \label{reward}
    R(\textbf{S}) = \mathcal{R} \cdot  e^{\alpha \cdot \sum_{t=1}^{T-1}r_t}
\end{equation}
where $\alpha$ is the parameter to balance the importance of user immediate rewards and we will conduct further analysis for this term in Section \ref{section:PA} to illustrate the relations.
The cumulative reward $e^{\sum_{t=1}^{T-1} r_t}$ is non-discounted and takes the exponential form as in analogy in supervised learning (i.e. binary cross-entropy) and RL-based solutions (i.e. TD3) where each step aims to learn the log scale policy output with the immediate reward.
As we will illustrate in the following sections, this design derives a flow matching objective with a simple immediate reward augmentation. The flow estimation process is detailed in Part (b) of Figure \ref{fig:framework}.



\subsubsection{Integrated Flow Matching}

Following the aforementioned reward integration, we propose to decompose the flow estimator of a given state $s_t$ into two corresponding components:
\begin{equation}
\label{eq:decompose}
\mathcal{F}(s_t) = \mathcal{F}_{R}(s_t) \cdot (\mathcal{F}_{I}(s_t))^\alpha
\end{equation}
where $\mathcal{F}_{R}(s_t)$ matches the flow of retention reward and $\mathcal{F}_{I}(s_t)$ matches the accumulated immediate rewards before step $t$:
\begin{equation}
\label{eq:immediate_reward}
    \mathcal{F}_{I}(s_t) = e^{\sum_{j=1}^{t-1} r_j}
\end{equation}
Different from $\mathcal{F}_R$ that needs to be optimized along with the GFN modules, $\mathcal{F}_I$ is a non-parametric function that only depends on the observed immediate reward.
Specifically for the terminal state, we have $\mathcal{F}_{R}(s_T)=\mathcal{R}$ and $\mathcal{F}_{I}(s_t)=e^{\sum_{j=1}^{T-1} r_j}$. 
Then, based on the flow matching objective, we can back-propagate the integrated reward towards intermediate steps:
\begin{equation} \label{eq:db_nonterminal}
    \mathcal{F}(s_t)\cdot P_F\left(s_{t+1}|s_t\right)=\mathcal{F}(s_{t+1}) \cdot P_B\left(s_t|s_{t+1}\right)
\end{equation}

Combining Eq. (\ref{eq:decompose}), Eq. (\ref{eq:immediate_reward}), and Eq. (\ref{eq:db_nonterminal}) together, we may derive the following simplified relation:

\begin{equation} \label{eq:non_terminal}
\begin{aligned}
    \mathcal{F}_{R}(s_t) 
    (\mathcal{F}_{I}(s_t))^{\alpha}P_F\left(a_t \mid s_t\right) 
    & = \mathcal{F}_{R}(s_{t+1}) (\mathcal{F}_{I}(s_{t+1}))^{\alpha} P_B\left(s_t|s_{t+1}\right) \\
    (e^{\sum_{j=1}^{t-1} r_j})^\alpha  \cdot \mathcal{F}_{R}(s_t) P_F\left(s_{t+1} \mid s_t\right) 
    & = (e^{\sum_{j=1}^{t} r_j})^\alpha  \cdot \mathcal{F}_{R}(s_{t+1}) P_B\left(s_t|s_{t+1}\right) \\
    \mathcal{F}_{R}(s_t) P_F\left(s_{t+}|s_t\right) 
    & =e^{\alpha r_t}  \cdot \mathcal{F}_{R}(s_{t+1}) P_B\left(s_t|s_{t+1}\right)
\end{aligned}
\end{equation}
The Eq. (\ref{eq:non_terminal}) suggests that within the realm of accurate model predictions, an increased action probability in $P_F$ correlates with either an enhanced immediate reward or a higher potential flow for future retention rewards.





\subsubsection{Integrated Detail Balance Objective}

Following the flow matching in Eq.\eqref{eq:non_terminal}, we derive the log-scale detailed balance learning objective as the following:
\begin{small}
\begin{equation}
    \mathcal{L}_{\mathrm{DB}}= 
    \begin{cases}
    \left( \log\mathcal{F}_{R}(s_t) + \log P_F\left(s_{t+1}|s_t\right) -\log \mathcal{F}_{R}(s_{t+1}) \right. \\
    \quad \left. -\log P_B\left(s_t|s_{t+1}\right) -\alpha \cdot r_t \right)^2 & 1 \leq t \leq T-1 \\
    \\
    \left(\log \mathcal{F}_R(s_t) -\log \mathcal{R}\right)^2 & t = T
    \end{cases}
\end{equation}
\end{small}
where each step's immediate reward $r_t$ appears and only appears in the corresponding step-wise DB loss, and the terminal state at $t=T$ does not observe an immediate reward and only matches the retention flow. This learning objective is further illustrated in Part (c) of Figure \ref{fig:framework}.
Systematically, our design of flow $\mathcal{F}_I$ for immediate rewards is non-parametric and can naturally integrate into the DB objective with a simple extra term.
In contrast, the retention reward requires the learning of $\mathcal{F}_R$ and the flow back-propagation with flow matching implicitly achieves the `retention attribution'.
The overall learning framework optimizes towards the integrated goal of $P(\mathbf{S})\propto \mathcal{F}(s_T) = \mathcal{F}_R(s_T)(\mathcal{F}_I(s_T))^\alpha$.
In general, we regard retention and immediate rewards as two complementary aspects of the user's impression of the recommendation policy. In our solution, immediate rewards provide direct guidance in each step, while the retention flow guides the policy with step-wise attribution.

Besides, the values of forward probability $P_F(\cdot)$ may approach zero deviating significantly from the valid region of the flow estimator in the log scale. 
This discrepancy can introduce high variance that may influence the stability of the gradient computation. 
Therefore we include a bias $\beta_F$ as the hyperparameter and the corresponding log scale estimation becomes $\log (P_F(\cdot) + \beta_f)$.
Similarly, we also include $\beta_B$ to stabilize the backward function learning and $\beta_r$ to reduce the reward variance.

\section{Experiment}
In this section, we present a comprehensive performance evaluation of the \name framework through experiments on a simulated user environment for two real-world datasets. Additionally, we extend our evaluation to include online A/B testing conducted on a commercial platform to validate \name's effectiveness in a live environment. The implementation details are provided in the Appendix \ref{Appendix:implementation}.

\subsection{Dataset}
We carried out our experiment using two real-world datasets.
\begin{itemize} [leftmargin=*]
    \item \textbf{Kuairand-Pure} \footnote{https://kuairand.com/} is an unbiased sequential recommendation dataset characterized by random video exposures. 

    \item \textbf{MovieLens-1M} \footnote{https://grouplens.org/datasets/movielens/1m/}, a widely-used benchmark for RSs, boasts a more extensive scale but with a sparser distribution. 
\end{itemize}

The KuaiRand dataset comprises 12 feedback signals, out of which we focus on six positive feedback signals: ‘click’, ‘view time’, ‘like’, ‘comment’, ‘follow’, and ‘forward’. We also consider two negative feedback signals: ‘hate’ and ‘leave’, due to their relevance. Feedback signals that occur less frequently are not included in our study to maintain analytical clarity. Additionally, we extend our analysis to the ML-1m dataset, a widely recognized benchmark in the field of Recommender Systems, which contains ratings from 6,014 users for 3,417 movies. For the ML-1m dataset, we classify movies rated above 3 as positive instances (indicative of a ‘like’) and the others as negative instances (indicative of a ‘hate’), thus enabling a nuanced understanding of user preferences. The statistics of datasets are presented in Table \ref{table:dataset}.

\begin{table}[h]
    \scriptsize
    \centering
    \small
    \caption{Datasets Statistics}
    \begin{tabular}{ccccc}
    \toprule
    Dataset & \# Users & \# Items & \# Interactions & \# Density \\
    \midrule
    Kuairand-Pure & 27,285 & 7,551	 & 1,436,609 & 0.70\% \\
    \midrule
    MovieLens-1M & 6,400 & 3,706 & 1,000,208 & 4.22\% \\
    \bottomrule
    \end{tabular}
    \label{table:dataset}
\end{table}

\subsection{Simulated User Environment}
We have chosen the KuaiSim retention simulator \cite{zhao2023kuaisim} for our study, which is designed to emulate long-term user behavior on short video recommendation platforms. KuaiSim features two key components: a leave module, responsible for predicting the likelihood of a user exiting a session and thereby ending an episode; and a return module, which estimates the daily probability of a user's return to the platform, expressed as a multinomial distribution.

\begin{table*}[h]
\centering
\caption{Overall Performance on two datasets for different models.}
\vspace{-3mm}
\begin{tabular}{@{}|l|l|cccccc|@{}}
\toprule
\multirow{2}{*}{Dataset} & \multirow{2}{*}{Metric} & \multicolumn{6}{c|}{Model} \\
\cmidrule(l){3-8}
 & & {TD3} & {SAC} & {DIN} & {CEM} & {RLUR} & {\name} \\
\midrule
\multirow{4}{*}{Kuairend-Pure} & Return Time & 2.382 & 2.373 & 1.947 & 1.889 & \underline{1.786} & \textbf{1.496*} \\
& Retention & 0.151 & 0.150 & 0.154 & 0.156 & \underline{0.159} & \textbf{0.163*}         \\
& Click Rate   & 0.800 & \underline{0.801} & 0.773 & 0.762 & 0.789 & \textbf{0.805}         \\
& Long View Rate  & 0.791 & \underline{\textbf{0.795}} & 0.764 & 0.757 & 0.778 & 0.794         \\
& Like Rate    & 0.852 & \underline{0.857} & 0.812 & 0.804 & 0.831 & \textbf{0.862}         \\
\midrule
\multirow{4}{*}{ML-1M} & Return Time    & 2.258 & 2.246 & 1.893 & 1.814 & \underline{1.723} & \textbf{1.479*}         \\
& Retention & 0.141 & 0.142 & 0.153 & 0.158 & \underline{0.160} & \textbf{0.165*}         \\
& Click Rate   & 0.461 & \underline{0.468} & 0.454 & 0.448 & 0.459 & \textbf{0.473}         \\
& Long View Rate  & 0.459 & \underline{0.463} & 0.455 & 0.453 & 0.457 & \textbf{0.464}         \\
& Like Rate   & 0.568 & \underline{0.571} & 0.541 & 0.524 & 0.561 & \textbf{0.574}         \\
\bottomrule
\end{tabular}
\label{table:overall}
\vspace{2mm}
\\``\textbf{{\Large *}}'': the statistically significant improvements (\ie two-sided t-test with $p<0.05$) over the best baseline.
\\ \underline{Underline}: the best baseline model. \textbf{Bold}: the best performance among all models. 
\end{table*}

\subsection{Baselines}
Our model is benchmarked against a diverse array of baselines, including several classic reinforcement learning methods and models specially designed for retention optimization:

\begin{itemize} [leftmargin=*]
    \item \textbf{CEM} \cite{deng2006cross}: Often employed as a surrogate in recommendation tasks, this method is known for its effectiveness and simplicity in various RL scenarios.
    \item \textbf{DIN} \cite{zhou2018deep}: This network uniquely features a local activation unit, enabling dynamic learning of user interest representations from historical behaviors.
    \item \textbf{TD3} \cite{fujimoto2018addressing}: Building on the foundation of DDPG, TD3 enhances performance through techniques like clipped double-Q learning, delayed policy updates, and target policy smoothing.
    \item \textbf{SAC} \cite{haarnoja2018soft}: An advanced off-policy RL algorithm adopts strategies including clipped double-Q learning for improved stability.
    \item \textbf{RLUR} \cite{cai2023reinforcing}: Specifically designed for enhancing long-term user engagement, this RL-based algorithm focuses on strategies that cater to sustained user interactions.
\end{itemize}

\subsection{Evaluation Metrics}
We assess the efficacy of \name using metrics that focus on both immediate user interactions and long-term retention:
\begin{itemize} [leftmargin=*]
    \item \textbf{Return Time:} Measuring the average duration between consecutive sessions, serving as a direct indicator of user retention.
    \item \textbf{Retention:} Capturing the average value of user retention reward as defined in Subsection \ref{section:PD}, reflecting long-term engagement.
    \item \textbf{Click Rate:} Calculating the rate of clicks per session, indicating immediate user engagement with the content.
    \item \textbf{Long View Rate:} Determining the likelihood of a session's view time exceeding 50\%, showcasing deeper content engagement.
    \item \textbf{Like Rate:} Assessing the rates of likes per session, further indicating user preference and satisfaction.
\end{itemize}
For a robust evaluation, we derive our final results by averaging the outcomes of the last 1000 episodes during the training phase.


\subsection{Overall Performance}
\begin{figure}[t]
    \centering    \includegraphics[width=0.98\linewidth]{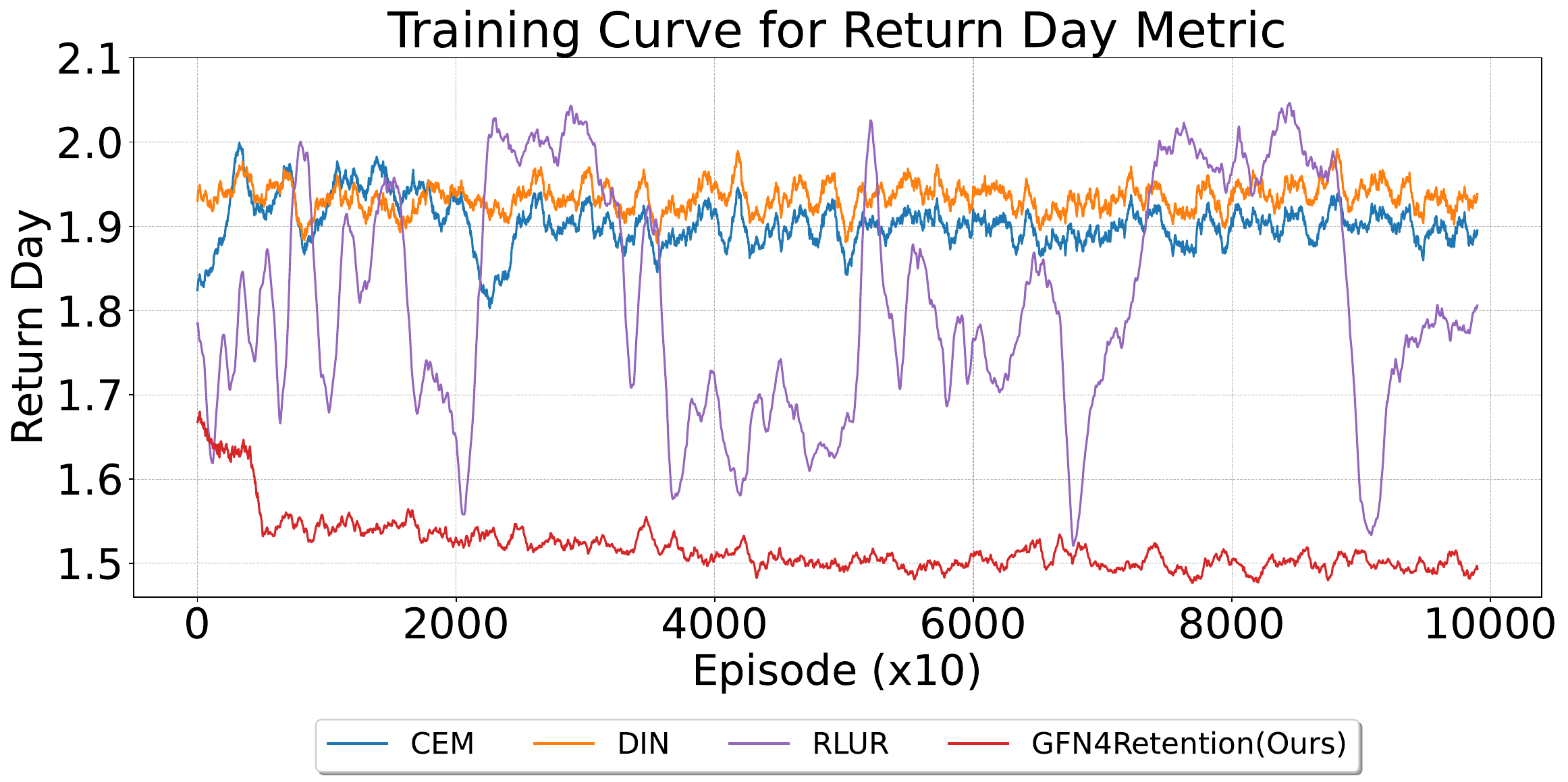}
    \vspace{-3mm}
    \caption{Training Curve for Return Day Metric.}
    \label{fig:TrainingCurve}
    \vspace{-4mm}
\end{figure}

To assess the effectiveness of our proposed GFN4Retention model, we conducted a comparative analysis of its overall performance against five baseline models on two datasets. The results are detailed in Table \ref{table:overall}. Additionally, we present the training curves of GFN4Retention alongside selected baselines focusing on the return day metric in Figure \ref{fig:TrainingCurve}. From these observations, we note that:
\begin{itemize} [leftmargin=*]
    \item The TD3 model registers the weakest performance in terms of retention metrics. It exhibits higher ‘return time' across both datasets, suggesting increased intervals between user sessions and, hence, lower retention rates. This model does not excel in any metric, likely due to its inadequate adaptation to shifts in the environment distribution and a policy weakly linked to specific user behavior patterns, resulting in suboptimal performance.
    \item Among all the baseline models, the RLUR model stands out in retention metrics and shows commendable results in optimizing immediate user feedback. This algorithm's design, which acknowledges the inherent biases of sequential recommendation tasks, adeptly captures user retention dynamics. Despite its strengths, the RLUR suffers from training volatility and requires more iterations to reach convergence.
    \item The SAC model emerges as the most proficient baseline in optimizing immediate user feedback. It boasts competitive performance across all metrics and leads in Long View Rate for the Kuairend-Pure dataset. Its approach to balancing expected returns with policy entropy enables effective modeling of user engagement.
    \item Our GFN4Retention model surpasses all other models, including the best baseline models, on several crucial metrics. It achieves the lowest ‘return time', indicating more frequent user engagement, and secures the highest scores in Retention and Like Rate, with statistically significant improvements. By integrating immediate feedback with the final retention signal in a meticulously structured manner, GFN4Retention boosts user retention while preserving the quality of immediate user feedback. The model's consistency and robustness are further evidenced by the most stable training curves among all baselines.
\end{itemize}

\noindent In summary, the GFN4Retention model demonstrates superior performance by effectively balancing immediate engagement with user retention, as indicated by its leading scores in critical metrics and significant improvements over the baseline models.

\subsection{Ablation Study}
To rigorously assess the contribution of individual modules within our proposed GFN4Retention model, we conducted an ablation study focusing on the context-detection module and the reward design. This study involves evaluating variations of the full model with the omission of particular settings to quantify their respective impacts. We describe the variant models as follows:

\begin{itemize} [leftmargin=*]
    \item \textbf{NCD (No Context Detection):} This model functions without the context-detection module, yet retains all other components, providing insight into the significance of context awareness in the user state encoder.

    \item \textbf{NIF (No Immediate Feedback):} In this variation, the reward structure is simplified to include only the retention reward, thereby isolating the effect of immediate feedback signals.

    \item \textbf{SIF (Simplified Immediate Feedback):} Here, we utilize a simplified immediate feedback reward approach, wherein session-wise accumulated rewards are considered solely at the terminal state, revealing the value of step-by-step reward accumulation.
\end{itemize}

\begin{figure}[t] 
    \centering
    \vspace{-1mm}{\subfigure{\includegraphics[width=0.325\linewidth]{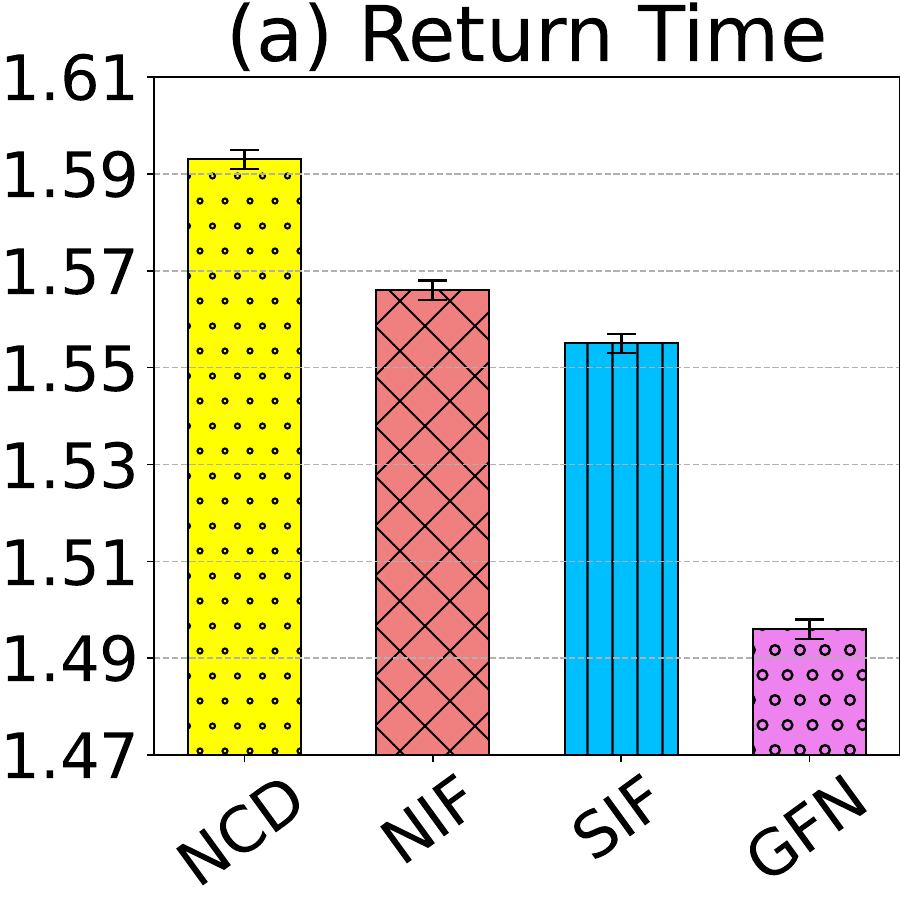}}}
	{\subfigure{\includegraphics[width=0.325\linewidth]{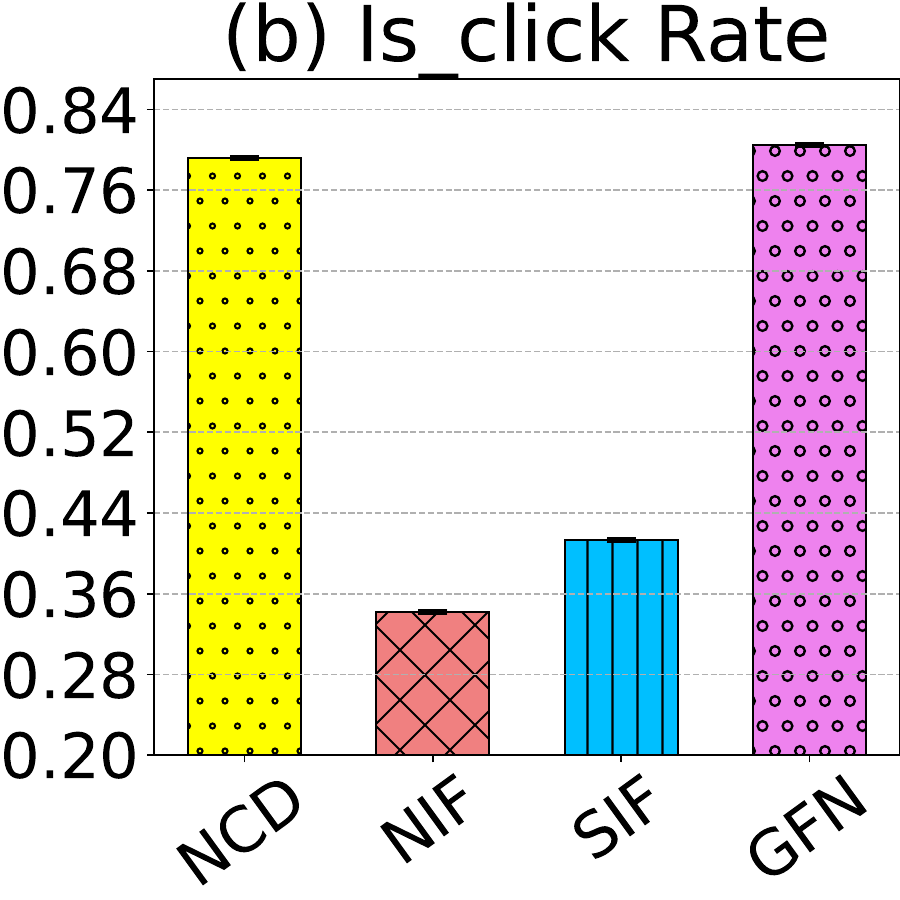}}}
	{\subfigure{\includegraphics[width=0.325\linewidth]{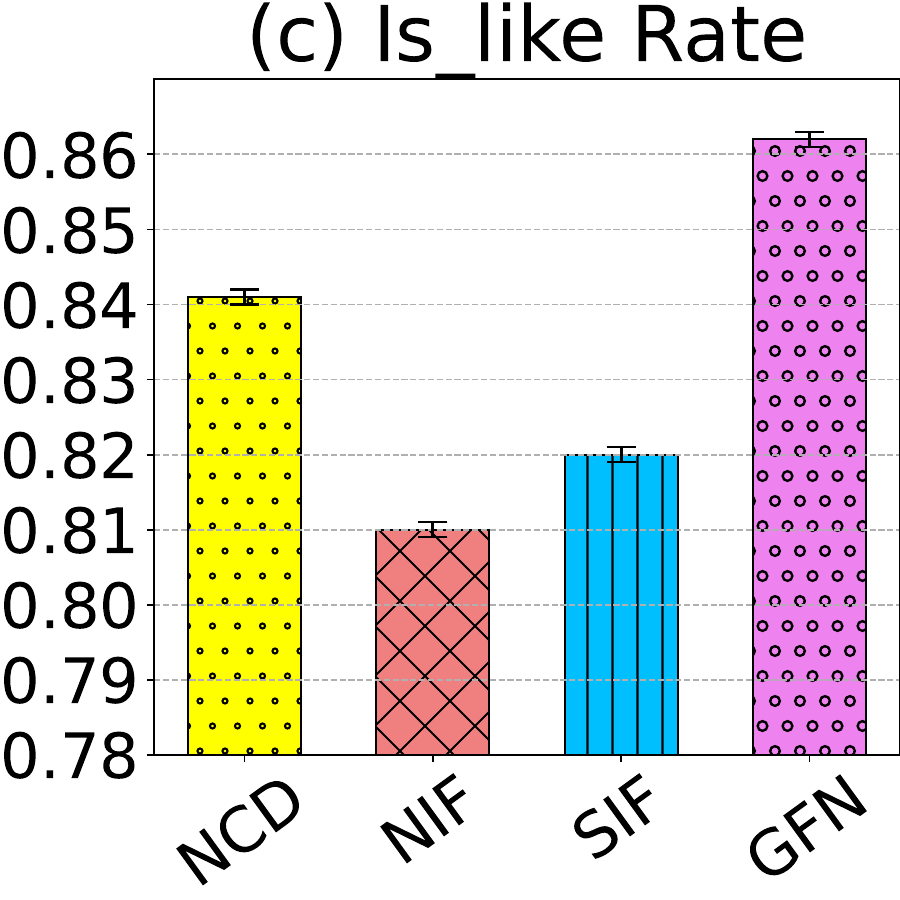}}}
    \vspace{-6mm}
    \caption{Ablation Study Results.}
    \label{Figure:ablation}
    \vspace{-3mm}
\end{figure}

\noindent The results of the ablation study on the GFN4Retention model, employing the Kuairand-Pure dataset, are depicted in Figure \ref{Figure:ablation}. This investigation yields several critical insights:

\begin{itemize} [leftmargin=*]
    \item The NCD variant demonstrates commendable performance in immediate feedback metrics such as ‘click rate' and ‘like rate'. However, it exhibits an approximate 6.7\% increase in ‘return time' relative to the comprehensive GFN4Retention model. This outcome can be principally ascribed to the absence of the context-detection module, which is pivotal in adaptively discerning user preferences that are integral to retention. These findings emphatically highlight the significance of the context-detection module within the model's architecture.
    
    \item In contrast to the complete GFN4Retention model, the NIF variant shows a marked 45\% reduction in ‘click rate' and a 5\% reduction in ‘like rate'. The lack of immediate feedback signals during the modeling process likely precipitates this decline. Additionally, there is a 4.7\% increase in ‘return time' compared to the GFN4Retention model, suggesting that only focusing on retention rewards is insufficient for capturing user preferences.

    \item The SIF variant, which incorporates immediate feedback signals into the terminal session reward, outperforms the NIF variant, registering gains in both short-term engagement metrics and user retention. This enhancement underscores the criticality of maintaining an equilibrium between immediate and sustained user engagement.

    \item The SIF variant exhibits a 40\% decrease in ‘click rate' and a 4\% decrease in ‘like rate', along with a 4\% increase in ‘return time', compared to the GFN4Retention model. These comparative metrics clearly demonstrate that our innovative reward structure, coupled with a refined detailed balance objective, significantly bolsters the model's proficiency in enhancing both immediate feedback and long-term user retention.
\end{itemize}

\subsection{Parameter Analysis} \label{section:PA}
\begin{figure}[t]
\label{PA}
    \centering    
    \vspace{-1mm}{\subfigure{\includegraphics[width=0.45\linewidth]{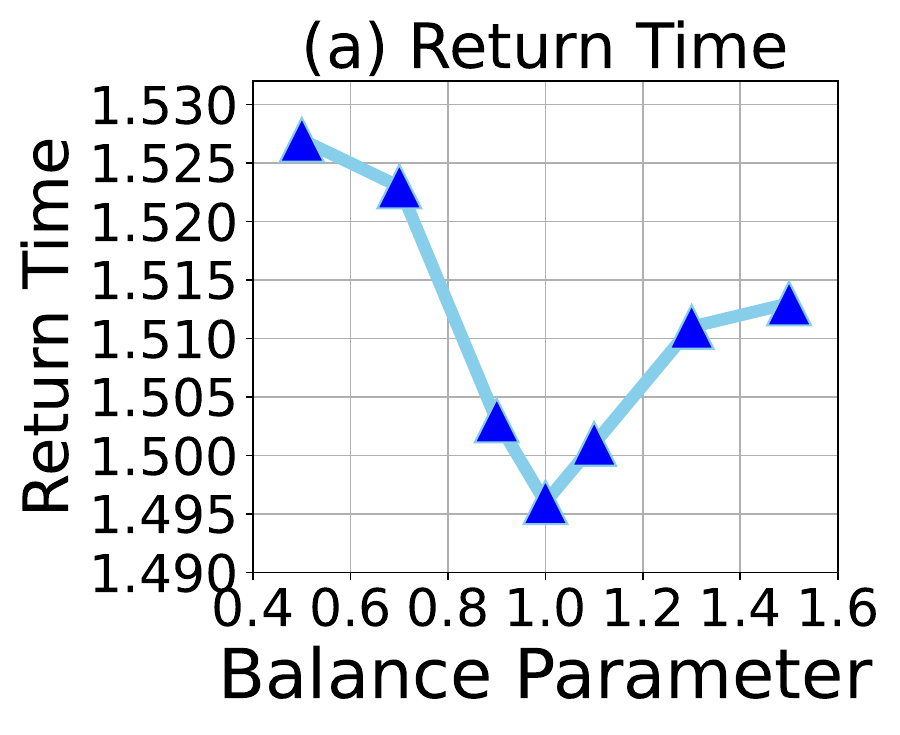}}}
	{\subfigure{\includegraphics[width=0.45\linewidth]{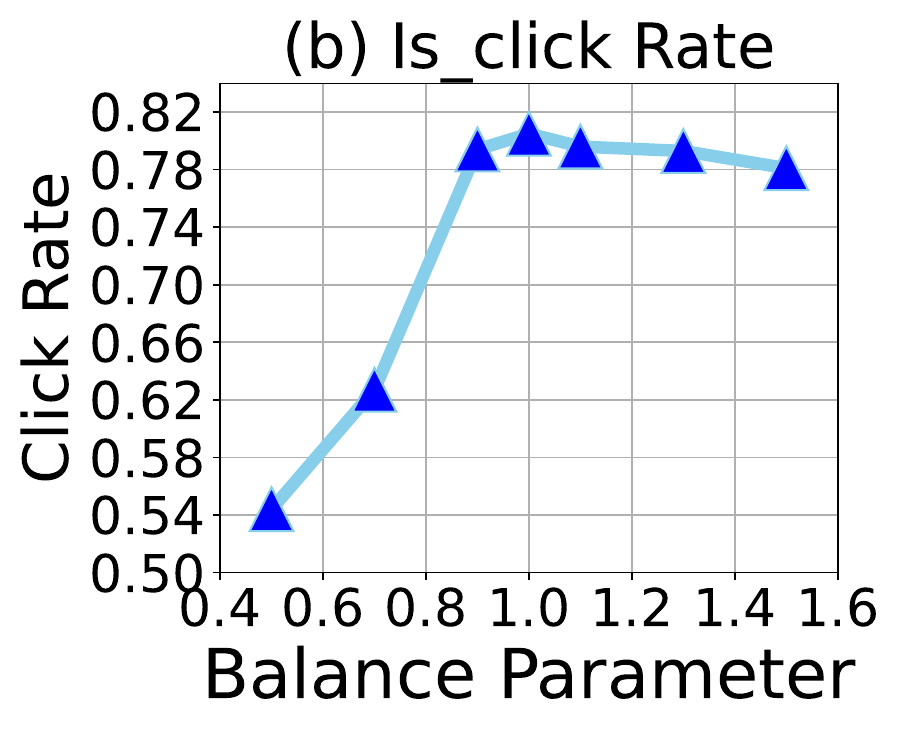}}}
    \vspace{-6mm}
    \caption{Balance Parameter Analysis.}
    \label{Figure:balance}
    \vspace{-3mm}
\end{figure}

To calibrate the significance of immediate user rewards within our model, we introduce the parameter $\alpha$ as part of the session-wise reward formulation, as delineated in Equation (\ref{reward}). This subsection is devoted to analyzing the impact of this balance parameter on model performance, particularly how the integration of immediate feedback signals affects outcomes. We assess the model's performance across a range of $\alpha$ values set at [0.5,0.7,0.9,1.0,1.1,1.3,1.5] with a focus on the ‘return time' and ‘click rate' metrics, the results of which are presented in Figure \ref{Figure:balance}. The following patterns emerge from the analysis:

\begin{itemize} [leftmargin=*]
    \item When $\alpha$ is reduced from 1.0 to 0.5, diminishing the weight of immediate user rewards, there is a marked decline in the ‘click rate'. This drop indicates that with less optimization towards immediate feedback, user engagement correspondingly decreases. Intriguingly, the ‘return time' metric also exhibits a significant increase, suggesting that the model's reduced focus on user engagement might adversely affect its capacity to capture user retention dynamics.
    \item Conversely, as $\alpha$ is incremented from 1.0 to 1.5, thereby increasing the emphasis on immediate rewards, there is a slight decrease in ‘click rate', particularly by 2\% when $\alpha$ reaches 1.5. This nuanced decline implies that an overly concentrated focus on immediate rewards could slightly detract from optimizing for engagement. Moreover, the heightened attention to immediate rewards seems to influence the retention optimization process, as evidenced by an increased ‘return time'.
\end{itemize}
These observations underscore the delicate balance required in tuning the $\alpha$ parameter to harmonize the dual objectives of optimizing the retention signal, thereby affirming the intricacies involved in reward structure design within our model.

\begin{figure}[h]
    \centering
    \includegraphics[width=\linewidth]{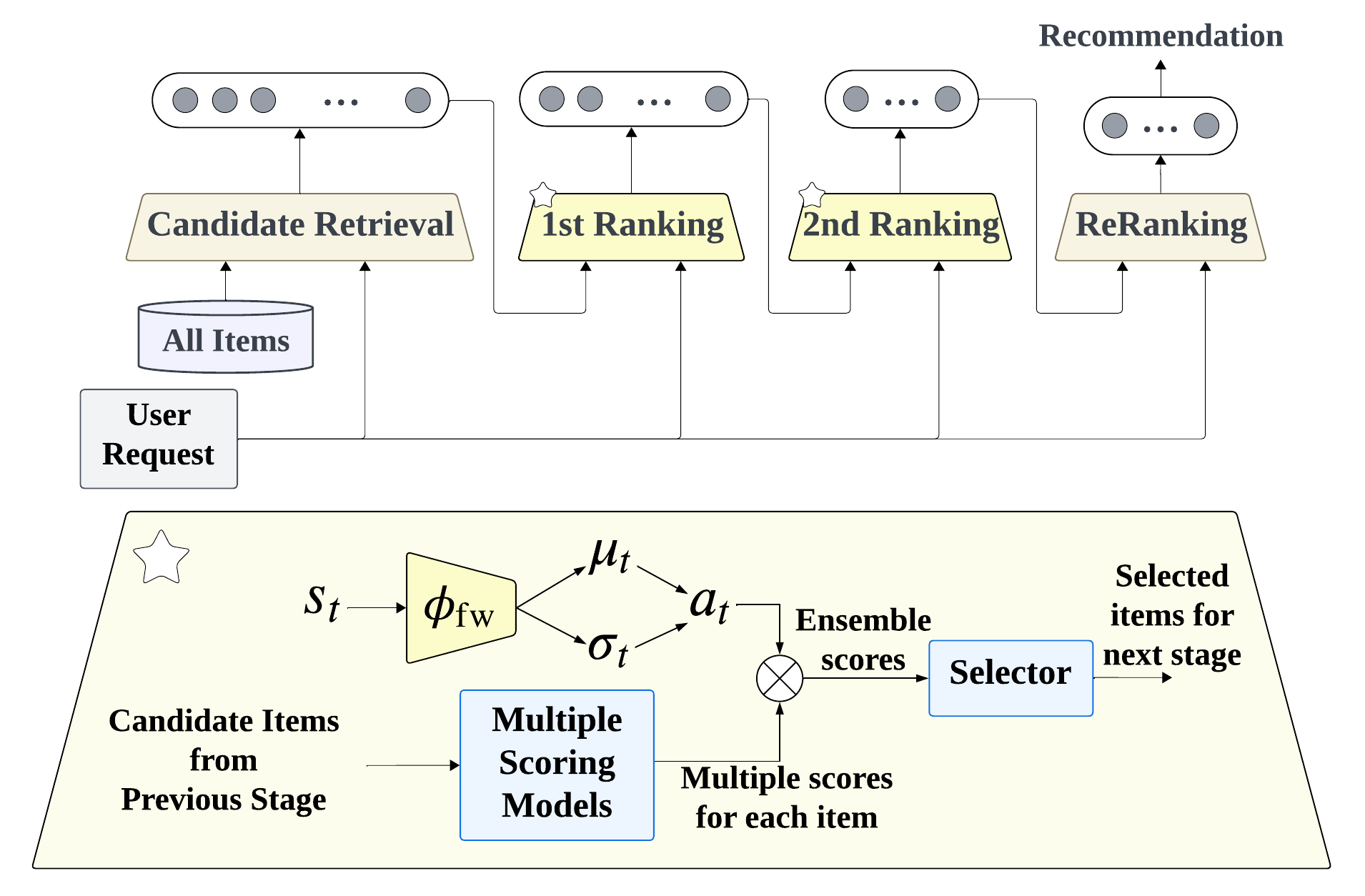}
    \vspace{-4mm}
    \caption{Deployment of \name in live experiments. 
    The bottom plot gives the detail of the two ranking stages where $\otimes$ represents the ensemble function that takes item scores from multiple ranking models as input and $a_t$ as weights of the scores.}
    \vspace{-4mm}
    \label{fig: online_deploy}
\end{figure}

\subsection{Live Experiment}

We conduct the live evaluation of our \name solution through an A/B test on a real-world industrial video recommendation platform.
The system serves billions of user requests every day and the daily candidate pool is around several million.
The overall recommender system consists of a relevant candidate retrieval stage and three rank-and-filter stages that gradually scale down the number of selected items before it is ready for exposure.
As shown in Figure \ref{fig: online_deploy}, we deploy \name in the ranking score ensemble modules of two of the ranking stages (i.e. the first ranking stage and the second-ranking stage) with separately learned policy models.
The output action corresponds to the weights that fuse the ranking scores from multiple score prediction models.
The baseline in the first ranking stage linearly combines the input ranking scores with empirically fixed parameters, while the baseline in the second ranking stage adopts an RL-based solution that automatically searches the action space.
For each experiment, we holdout 10\% of the total online traffic for the \name solution and 20\% of the total online traffic for the baseline.
For the retention signal, we use the reciprocal return time gap as the retention reward of a user session and the normalized watch time as an immediate reward.
Correspondingly, the evaluation consists of the next-day user return frequency and the average watch time which are evaluated on daily basis. Both metrics are better if larger.
We summarize the results in Table \ref{tab: live_result} which proves that the \name method can significantly improve the user's retention and the corresponding immediate reward.
When focusing on the target user group with relatively lower activity in the system, the improvement in retention is more significant.
Note that we have applied the proposed method in the two major ranking scenarios in the system, indicating its generalization ability and scalability in different stages.

\textcolor{red}{
\begin{table}[h]
    \centering
    \caption{Online A/B Test Performance.}
    \begin{tabular}{c|c|cc}
        \toprule
        & $1^\mathrm{st}$ Stage & $2^\mathrm{nd}$ Stage \\
        \midrule
        overall next-day retention & +0.015\%* & +0.002\% \\
        target users' next-day retention & +0.069\%* & +0.056\%* \\
        \midrule
        watch\_time & +0.558\%* & +0.224\%*\\
        \bottomrule
    \end{tabular}
    \label{tab: live_result}
    \vspace{2mm}
    \\``\textbf{{\Large *}}'': the statistically significant improvements over the baseline.
\end{table}
}
\section{Related Work}
In this section, we briefly discuss existing research related to RL-based recommender systems and retention optimization.

\subsection{\textbf{Reinforcement Learning Based RSs}}

The application of Reinforcement Learning (RL) for recommendations is justified by the underlying Markov Decision Process framework, which is fundamental to the RL paradigm~\cite{sutton2018reinforcement,afsar2021reinforcement,wang2022surrogate,zhang2022multi, shani2005mdp}. RL's primary benefit in this context is its focus on maximizing the expected cumulative reward from user interactions over time, rather than just enhancing immediate recommendations. In environments with limited recommendation options, methods that utilize tabular approaches ~\cite{mahmood2007learning,moling2012optimal} or value function estimation ~\cite{taghipour2007usage,zheng2018drn,zhao2018recommendations,ie2019slateq} have been employed to assess the long-term implications of recommendations. Conversely, in situations characterized by vast action spaces, policy gradient ~\cite{sun2018conversational,chen2019top,chen2019large}, and actor-critic methodologies ~\cite{sutton1999policy,peters2008natural,bhatnagar2007incremental,degris2012model,dulac2015deep,liu2018deep,liu2020state,xin2020self, zhao2020whole,cai2023two} are preferred for their ability to steer the recommendation policy toward higher-quality outcomes. The complexity of optimizing for multiple metrics is addressed in the literature on multi-objective optimization, highlighting the varying behavioral patterns among users \cite{chen2021generative,cai2023two}. To bridge the discrepancy between real-world user interactions and offline assessments, user simulators have become a pivotal tool for researchers~\cite{ie2019recsim,zhao2019deep,zhao2023kuaisim}. Our methodology extends this direction by performing offline training in simulated cross-session environments for user retention optimization.

In parallel, Generative Flow Networks (GFNs) have surfaced as a groundbreaking approach \cite{bengio2021flow}, drawing parallels with RL but pushing the boundaries in terms of generating diverse, high-quality samples from intricate distributions. Notably, GFNs have demonstrated their efficacy in tackling list-wise recommendation challenges \cite{liu2023generative}, showcasing their utility in recommendation systems. 
Our research leverages GFNs with novel flow-matching formulations and tailored objective functions to refine user retention optimization.
We notice that the generative process of GFN may remind readers about the diffusion-based methods which have also been studied in sequential recommendations \cite{li2023diffurec,yang2024generate}.
However, they are not designed for sparse and delayed return signal modeling since they require precise and abundant representations for unsupervised learning. Intuitively, we believe our GFN-based solution fits better to this problem since it is an energy-based model that can predict the delayed retention signal while following the iterative recommendation process. 


\subsection{Retention Optimization for RSs}
Shifting away from the traditional focus on immediate feedback, the research community has started delving into strategies for enhancing users' long-term satisfaction \cite{wu2017returning,cai2020predicting, carvalho2013users,chen2019serendipity}. Along this direction, several studies have investigated methods to boost long-term user engagement by analyzing metrics like dwell time \cite{kapoor2014hazard,chandar2022using,zou2019reinforcement}. However, the domain of user retention-oriented optimization remains relatively underexplored. Notably, a few pioneering efforts have aimed to predict user retention through innovative perspective \cite{cai2023reinforcing,zhao2023user,ding2023interpretable,gwalani2022studying}. , such as the rationale contrastive multi-instance learning approach, designed to elucidate the factors influencing user retention and thereby augment its interpretability \cite{ding2023interpretable}.

A notable advancement includes leveraging decision transformer-based models to tackle user retention challenges, and ingeniously reframing the reinforcement learning (RL) problem as an autoregression issue \cite{zhao2023user}. Furthermore, some researchers argue that user retention—viewed as feedback accumulated over multiple interactions with the system—presents a complex challenge in attributing retention rewards to individual items or sequences \cite{cai2023reinforcing}. They propose conceptualizing this issue using reinforcement learning to minimize the cumulative time intervals across sessions. Our work builds upon these insights into retention optimization, uniquely considering the integration of immediate feedback within the modeling framework to better capture the nuances of user retention dynamics. Moreover, our approach has demonstrated efficacy across both offline datasets and large-scale online platforms.

\section{Conclusion}
In this work, we delve into optimizing user retention within recommender systems, a critical aspect for fostering sustained user engagement. Recognizing the intricate nature of long-term user interactions, we conceptualize the retention signal as a holistic measure of user satisfaction at the session's conclusion. Our approach models this comprehensive estimation through a generative flow, ingeniously back-propagating the retention reward to each user's immediate feedback within a session. By employing a simplified flow matching technique alongside a novel DB loss function, our model optimizes long-term retention while also integrating immediate feedback signals. The efficacy of our methodology is rigorously validated through extensive offline empirical evaluations on publicly available datasets and real-world online A/B testing on a commercial platform, demonstrating its practical applicability and effectiveness in enhancing user retention.

\section*{ACKNOWLEDGEMENTS}
This research was partially supported by Kuaishou, Research Impact Fund (No.R1015-23), APRC - CityU New Research Initiatives (No.9610565, Start-up Grant for New Faculty of CityU), CityU - HKIDS Early Career Research Grant (No.9360163), Hong Kong ITC Innovation and Technology Fund Midstream Research Programme for Universities Project (No.ITS/034/22MS), Hong Kong Environmental and Conservation Fund (No.88/2022), and SIRG - CityU Strategic Interdisciplinary Research Grant (No.7020046, No.7020074).

\begin{appendix}



\section{IMPLEMENTATION DETAILS} \label{Appendix:implementation}
In our study, we implement a cross-session environment setup using the KuaiSim Simulator to assess our models. Here are the detailed configurations for each dataset:

\noindent \textbf{For the Kuairand-Pure Dataset:}
\begin{itemize} [leftmargin=*]
    \item We utilize a CrossSessionBuffer of size 100,000.
    \item The policy model is configured with a Transformer architecture, featuring 4 attention heads and an embedding size of 32.
    \item Learning rates are set to 0.00002 for the flow estimator and 0.0001 for both the forward and backward estimators.
    \item All estimators share an embedding hidden dimension of 128.
    \item Optimization is conducted using the Adam optimizer, with a batch size of 128.
    \item A default balance parameter $\alpha$ of 1 is applied.
    \item Smooth offset values are set at 1.0 for both forward $\beta_F$ and backward $\beta_B$ probabilities, with a reward smooth offset $\beta_r$ of 0.6.
\end{itemize}

\noindent \textbf{For the MovieLens-1M Dataset:}
\begin{itemize} [leftmargin=*]
    \item We utilize a CrossSessionBuffer of size 10,000.
    \item The policy model is configured with a Transformer architecture, featuring 4 attention heads and an embedding size of 32.
    \item Learning rates are set to 0.0003 for the flow estimator and 0.001 for both the forward and backward estimators.
    \item All estimators share an embedding hidden dimension of 64.
    \item Optimization is conducted using the Adam optimizer, with a batch size of 32.
    \item A default balance parameter $\alpha$ of 1 is applied.
    \item Smooth offset values are set at 0.8 for both forward $\beta_F$ and backward $\beta_B$ probabilities, with a reward smooth offset $\beta_r$ of 0.6.
\end{itemize}

\noindent Both datasets undergo 100,000 training steps to ensure comprehensive model evaluation.

\section{ALGORITHM} \label{Appendix:algorithm}
In this section, we elucidate the optimization algorithm underpinning our model, presented through explanatory pseudo-code. As delineated in Algorithm 1, the training procedure for GFN4Retention is methodically straightforward, adhering to a clearly defined sequence of operations that ensure the model's convergence to the desired objectives. This structured approach facilitates ease of replication and verification of the results reported herein.

\begin{algorithm}
\caption{Training Process for GFN4Retention}
\begin{algorithmic}[1]
\STATE \textbf{\# Apply current policy in running episodes:}
\STATE \textbf{procedure} ONLINE INFERENCE
\STATE Initialize replay buffer \(\mathcal{B}\).
\WHILE{True, in each running episode do}
    \STATE Observe user request \( \textbf{S}_{u,0} \).
    \STATE Initial the item sequence \( s^0 \leftarrow \emptyset \)
    \FOR{\( t \in \{1, \ldots, T\} \) do}
        \STATE Sample item list \( a_t \sim P(\cdot|u, s^{t-1}) \) with current policy.
    \ENDFOR
    \STATE Obtain user responses \( \textbf{y} \) from online environment and calculate the retention reward \( \mathcal{R}(u, s) \).
    \STATE \( (u, s, \mathcal{R}(u, s), \textbf{y}) \rightarrow \mathcal{B} \)
\ENDWHILE
\STATE \textbf{end procedure}
\STATE \textbf{\# Simultaneous training on the buffer:}
\STATE \textbf{procedure} TRAINING
\STATE Initialize all trainable parameters in the policy (e.g., \( \phi \), $\theta_1$ and $\theta_2$) in GFN4Retention)
\STATE Wait until \(\mathcal{B}\) has stored minimum amount of data points.
\WHILE{Not Converged, in each iteration do}
    \STATE Obtain mini-batch sample \( (u, s, R_l(u, s), \textbf{y}) \rightarrow \mathcal{B} \).
    \STATE Calculate flow estimator $\mathcal{F}(s^t;\phi)$, forward estimator $P_F(a_t|u, s^{t-1};\theta_1)$ and backward estimator $P_B(s_{t-1}|u, s^t;\theta_2)$ for each generation step \( t \).
    \STATE Update the policy through one step of gradient descent using Adam on the DB loss function \( \mathcal{L}_{DB} \).
\ENDWHILE
\STATE \textbf{end procedure}
\end{algorithmic}
\end{algorithm}

\begin{table*}
\centering
\caption{Hyperparameter Settings and Choices}
\label{tab:hyperparameters}
\begin{tabular}{@{}l|l|l|l@{}}
\toprule
Dataset         & Hyper-parameter      & Tuning Range    & Our Choice \\ \midrule
                & Number of heads & [2, 4, 6, 8]      & 4          \\
                & Embedding size & [8, 16, 32, 64, 128]      & 32         \\
                & LR for flow estimator  & [0.00001, 0.00002, 0.0001, 0.0002] & 0.00002    \\
KuaiRand-Pure   & LR for forward estimator & [0.00005, 0.0001, 0.0002, 0.001]& 0.0001     \\
                & Batch size    & [32, 64, 128, 256]       & 128        \\
                & Forward offset $\beta_F$ & [0.2, 0.4, 0.6, 0.8, 1.0, 1.2]      & 1          \\
                & Backward offset $\beta_B$ & [0.2, 0.4, 0.6, 0.8, 1.0, 1.2]    & 1          \\
                & Reward offset $\beta_r$  & [0.2, 0.4, 0.6, 0.8, 1.0, 1.2]     & 0.5        \\ \midrule
                & LR for flow estimator  & [0.0001, 0.0002, 0.0003, 0.001] & 0.0003     \\
                & LR for forward estimator & [0.0001, 0.0002, 0.001, 0.002]& 0.001      \\
MovieLens-1M    & Batch size    & [32,64,128,256]       & 64         \\
                & Forward offset $\beta_F$ & [0.2, 0.4, 0.6, 0.8, 1.0, 1.2]      & 0.8        \\
                & Backward offset $\beta_B$  & [0.2, 0.4, 0.6, 0.8, 1.0, 1.2]    & 0.8        \\
                & Reward offset $\beta_r$  & [0.2, 0.4, 0.6, 0.8, 1.0, 1.2]      & 0.5        \\ \bottomrule
\end{tabular}
\end{table*}

\section{HYPER-PARAMETER SELECTION} \label{Appendix:parameter}
In our study, we meticulously tuned the hyperparameters of the \name model to reinforce the reliability of our experimental outcomes. The summarized performance metrics, spanning two datasets, are documented for detailed examination in Table \ref{tab:hyperparameters}.

\section{Probablistic Flow in Continuous Space}
\label{Appendix:continuous}


For a given state, the flow estimator \( \mathcal{F}(s_t) \) represents the generation likelihood of reaching that particular state.
Without loss of generality, assume each sample consists of the transition information between the current state $s_t=x_1$ and the next state $s_{t+1}=y$.
The sampled action is denoted as $a_t=z$.
Our forward function is associated with a Gaussian actor $\mu,\sigma=\phi_\mathrm{fw}$ where corresponding action distribution follows $a_t\sim \mathcal{N}(\mu,\sigma) \mathcal{F}(s_t=x_1)$.
Then the forward function for the observed transition is $P_F(s_{t+1}|s_t)=P(a_t=z)$
The flow matching property holds at this observed point with the joint likelihood 
\begin{equation}
\begin{aligned}
    & P_F(s_{t+1}=y|s_t=x_1)\mathcal{F}(s_t=x_1)\\
    = & P(s_t=x_1,s_{t+1}=y)\\
    = & P_B(s_t=x_1|s_{t+1}=y)\mathcal{F}(s_{t+1}=y)
\end{aligned}
\end{equation}
Additionally, the flow of the next state essentially expresses the expected reachability in the generation process:
\begin{equation}
\mathcal{F}(s_{t+1}) = \int_{s_t} P_F(s_{t+1}|s_t) \mathcal{F}(s_t)
\end{equation}
which addresses the infinite number of previous states for the given next state.
Note that different from the forward policy that follows the Gaussian distribution, the distribution of state flow $\mathcal{F}$ and the backward flow $P_B$ could be complicated.
So we use neural networks to approximate these likelihoods in our design.


\end{appendix}
 \twocolumn

\bibliographystyle{ACM-Reference-Format}
\balance
\bibliography{ref}

\end{document}